\documentclass[a4paper,10pt]{article}

\newcommand{\nin}{\noindent}
\newcommand{\be}{\begin{equation}}
\newcommand{\ee}{\end{equation}}
\newcommand{\bea}{\begin{eqnarray}}
\newcommand{\eea}{\end{eqnarray}}

\newcommand{\hf}{\frac{1}{2}}
\newcommand{\nn}{\nonumber\\}

\begin{document}

\title{A misleading Wilsonian fixed point}

\author{Jean Alexandre\\ Department of Physics, King's College London, WC2R 2LS, UK\\ jean.alexandre@kcl.ac.uk}

\maketitle

\begin{abstract}

\nin We exhibit here, for a scalar theory, an apparently non-trivial Wilsonian fixed point, which surprisingly 
describes a free theory. This modest note is an observation which can be of interest in the framework of
functional methods in Quantum Field Theory.

\end{abstract}

\vspace{.5cm}

The search for Wilsonian fixed points is important in Quantum Field Theory, as these represent configurations
independent of the energy scale of the process under study, and hence for which quantum fluctuations  
are controlled.
We exhibit here, for a scalar theory, an exact solution of a renormalisation group equation, leading to  
an IR fixed point configuration, which is not analytical in the dynamical field. Although this configuration seems 
to describe an interacting theory, its non-analiticity 
prevents from defining a loop expansion. We show then that the path integral quantization of an UV configuration similar to the latter fixed point can be recovered by the quantization of a free
theory, by implementing a functional change of variable in the partition function.
The non-analytic and apparently non-trivial fixed point is therefore just another representation of a free theory, 
which explains why it satisfies an exact renormalization equation.

The Jacobian generated by the functional change of variable is essential for the equivalence between the 
would-be non-trivial fixed point and a free theory. More generally, the definition of the integration measure 
can play an important role in the functional integral, 
especially for gauge theories defined on curved space times \cite{curved}.
Also, the Jacobian arising from a Weyl transformation of the world sheet metric, in the framework of non-critical 
strings, it responsible for the generation of the Liouville action \cite{dk}. Finally, it has been shown in the framework of
scalar electrodymanics, that the parametrization of matter using polar coordinates, in the field space,
also leads to a non-trivial quantization \cite{toms}.

{\it Wegner Houghton exact renormalization equation.}
Consider a scalar field $\Phi=\phi+\tilde\phi$, where the IR field $\phi$ has non-vanishing Fourier components for 
momenta smaller than some value $k$, and the UV field $\tilde\phi$ has non-vanishing Fourier components for 
momenta $p$ such that $k<|p|\le\Lambda$, where $\Lambda$ is a fixed cut off.
The bare Euclidean action we consider is, in $D$ dimensions, 
\be\label{bareaction}
S_\Lambda[\Phi]=\int d^Dx\left\lbrace \hf Z_\Lambda(\Phi)\partial_\mu\Phi\partial^\mu\Phi+U_\Lambda(\Phi)\right\rbrace .
\ee
The Wilsonian effective action for the low energy field $\phi$ is defined by the functional 
integration over the high energy Fourier modes of $\tilde\phi$:
\bea
S_k[\phi]=\int{\cal D}[\tilde\phi]\exp\left( -S_\Lambda[\phi+\tilde\phi]\right).
\eea
We assume then, in the framework of the gradient expansion, that
the Wilsonian effective action can be truncated to the following functional dependence, for any value of $k$,
\be\label{action}
S_k=\int d^Dx\left\lbrace \hf Z_k(\phi)\partial_\mu\phi\partial^\mu\phi+U_k(\phi)\right\rbrace ,
\ee
where higher derivative operators are neglected, as they become negligible in the IR.
It can then be shown \cite{old}, following the Wegner-Houghton procedure \cite{WH}, 
that an infinitesimal step $k\to k-dk$ leads, in the limit $dk\to 0$, to the following renormalization equation for the potential
\be\label{wh}
\partial_k U_k(\phi)=\partial_k U_k(\phi_0)
-\frac{\Omega_D k^{D-1}}{2(2\pi)^D}\ln\left(\frac{Z_k(\phi)k^2+U_k^{''}(\phi)}{Z_k(\phi_0)k^2+U_k^{''}(\phi_0)}\right),
\ee
where $\Omega_D$ is the solid angle in dimension $D$, and $\phi_0$ is a fixed value.
In the framework of the assumption (\ref{action}), the equation (\ref{wh}) is exact: it is a self consistent equation and 
corresponds to a ressumation of all the loops.
It gives the evolution for the potential only: the Wegner-Houghton procedure can't 
generate the flow for the wave function renormalization $Z_k(\phi)$, due to singularities generated by the sharp 
cut off, in the evolution of derivative operators. In order to obtain the evolution for derivative operators
with $k$, one needs to introduce a smooth cut off, in the spirit of the Polchinski equation \cite{polchinski},
or using the average action formalism \cite{smooth}.
The evolution for $Z_k(\phi)$ with $k$ would then depend on the precise cut off function which is chosen, but
not its IR limit $k\to 0$. In this note we will concentrate on the
evolution equation for the potential, which is independent of the cut off function used.

{\it Non-trivial Wilsonian fixed point?}
It is interesting to see that the evolution equation (\ref{wh}) has a non-trivial exact solution, which is
non-analytic in $\phi$, and reads
\bea\label{solution}
Z_k(\phi)&=&z_k\left(\frac{\phi_0}{\phi}\right)^2\nn
U_k(\phi)&=&\left(m^D+\frac{\Omega_Dk^D}{D(2\pi)^D}\right)\ln\left(\frac{\phi}{\phi_0}\right),
\eea
where $m$ is a constant mass and $z_k$ is a dimensionless function of $k$, which could be determined 
if we used a smooth cut off procedure.
The solution (\ref{solution}) leads to the following IR fixed point
\bea\label{fixedpoint}
Z^\star(\phi)&=&z_0\left( \frac{\phi_0}{\phi}\right)^2\nn
U^\star(\phi)&=&m^D\ln\left(\frac{\phi}{\phi_0}\right),
\eea 
which seems to describe a
non-trivial theory, as an expansion around $\phi=\phi_0$ would lead to a whole series of 
power and derivative interactions.
But we show now that it actually represents a free theory, which explains that it 
is a Wilsonian fixed point.

{\it Equivalence with a free theory.}
The fixed point (\ref{fixedpoint}) can be understood with the following argument.
Let $\psi$ be a massless scalar field, for which the partition function for the free theory in Euclidean space is 
\be\label{partition}
{\cal Z}=\int{\cal D}[\psi]\exp\left(-\hf\int d^Dx ~\partial_\mu\psi\partial^\mu\psi \right).
\ee
It is assumed that the Fourier components of $\psi$ vanish above a given cut off $\Lambda$, 
which is therefore the only dimensionful parameter in the problem.
We make a change of variable in the functional integral and write 
\be\label{change}
\psi=\Lambda^{D/2-1}\ln\left(\frac{\phi}{\Lambda^{D/2-1}}\right),
\ee
where $\phi$ is a scalar field with positive values. 
The action is then, in terms of $\phi$,
\be
\hf\int d^Dx~\partial_\mu\psi\partial^\mu\psi
=\frac{\Lambda^{D-2}}{2}\int d^Dx~\frac{\partial_\mu\phi\partial^\mu\phi}{\phi^2}.
\ee
To the change of functional variable (\ref{change}) corresponds a non-trivial Jacobian, such that 
the measure of integration becomes
\bea
{\cal D}[\psi]&=&\Pi_x d\psi(x)\nn
&=&\Pi_x\left(\Lambda^{D/2-1}\frac{d\phi(x)}{\phi(x)}\right)\nn
&=&\Big(\Pi_xd\phi(x)\Big)\Pi_x\exp\Big(-\ln\left(\phi(x)/\Lambda^{D/2-1}\right)\Big)\nn
&=&{\cal D}[\phi]\exp\Big(-\Sigma_x\ln\left(\phi(x)/\Lambda^{D/2-1}\right)\Big).
\eea
In the continuous limit, the summation over space time lattice sites gives
\be
\Sigma_x\to a\Lambda^D\int d^Dx,
\ee 
where $a>0$ is a dimensionless constant, such that we obtain
\be
{\cal D}[\psi]={\cal D}[\phi]\exp\left(-a\Lambda^D\int d^Dx~\ln\left(\frac{\phi(x)}{\Lambda^{D/2-1}}\right)\right).
\ee
The partition function (\ref{partition}) finally reads $
{\cal Z}=\int{\cal D}[\phi]\exp\left(-S_{eff}[\phi]\right)$, where the effective action is
\be\label{Seff}
S_{eff}[\phi]=\Lambda^D\int d^Dx\left\lbrace 
\frac{\Lambda^{-2}}{2}\frac{\partial_\mu\phi\partial^\mu\phi}{\phi^2}
+a\ln\left(\frac{\phi(x)}{\Lambda^{D/2-1}}\right)\right\rbrace .
\ee
We can see that the fixed point (\ref{fixedpoint}) is recovered, showing that the latter represents a free theory.
Several concluding remarks can be done:

\begin{itemize}

\item In order for the change of variable (\ref{change}) to lead to the action (\ref{Seff}),
one needs to consider a Euclidean metric, which is a natural framework for Wilsonian transformations;

\item The configuration (\ref{fixedpoint})
cannot be truncated with the criterium of relevance/irrelevance of 
field operators, as the free theory could not be recovered by implementing the inverse change of variable 
\be
\phi=\Lambda^{D/2-1}\exp(\Lambda^{1-D/2}\psi).
\ee

\item A configuration of time-dependent bosonic string in graviton and dilaton backgrounds, similar to the fixed point 
(\ref{fixedpoint}), was found in \cite{AEM1}, and satisfies conformal invariance non-perturbatively
in $\alpha^{'}$. In this case, 
though, the corresponding action does not represent a free string, as a consequence of the interaction between
the different fields present in the theory. Indeed, the non-trivial action which was found is
\be
S=\frac{1}{4\pi\alpha^{'}}\int d^2\xi\sqrt\gamma\left\lbrace 
\gamma^{ab}\frac{A}{(X^0)^2}\eta_{\mu\nu}\partial_a X^\mu\partial_b X^\nu
+\alpha^{'}R^{(2)}\phi_0\ln X^0\right\rbrace ,
\ee
where $A$ and $\phi_0$ are constants,
$\gamma_{ab}$ is the world sheet metric, $R^{(2)}$ is the corresponding curvature scalar,
 and $X^\mu$ are the
dynamical fields living on the world sheet. 
The change of variable $X^0\to \exp(X^0)$ leads to the
following effective action, after taking into account the Jacobian of the transformation in the Poliakov
path integral,
\bea
S_{eff}&=&\frac{1}{4\pi\alpha^{'}}\int d^2\xi\sqrt\gamma\Big\{\gamma^{ab}A~\eta_{00}\partial_a X^0\partial_b X^0\\
&& +\gamma^{ab}A~\eta_{ij}~e^{-2X^0}\partial_a X^i\partial_b X^j+\alpha^{'}R^{(2)}\phi_1 X^0\Big\},\nonumber
\eea
where $\phi_1$ is a constant. The corresponding target space is therefore described (in the string frame)
by a de Sitter metric.

\end{itemize}


\begin{thebibliography}{99}

\bibitem{curved} R.~K.~Unz, 
  Nuovo Cim. A{\bf 92}: 397 (1986)

\bibitem{dk} J.~Distler and H.~Kawai
  Nucl.\ Phys.\ B{\bf 321} (1989) 509.

\bibitem{toms}
I.~H.~Russell and D.~J.~Toms,
  Phys.\ Rev.\  D {\bf 39} (1989) 1735.

\bibitem{old}J.~Alexandre, V.~Branchina and J.~Polonyi,
  Phys.\ Rev.\  D {\bf 58} (1998) 016002
  [arXiv:hep-th/9712147].

\bibitem{WH}F.~J.~Wegner and A.~Houghton,
  Phys.\ Rev.\  A {\bf 8} (1973) 401.

\bibitem{polchinski}J.~Polchinski,
  Nucl.\ Phys.\  B {\bf 231} (1984) 269.

\bibitem{smooth}For a review, see for example J.~Berges, N.~Tetradis and C.~Wetterich,
  Phys.\ Rept.\  {\bf 363} (2002) 223
  [arXiv:hep-ph/0005122], and references therein.

\bibitem{AEM1}J.~Alexandre, J.~R.~Ellis and N.~E.~Mavromatos,
  JHEP {\bf 0612} (2006) 071
  [arXiv:hep-th/0610072].


\end{thebibliography}
\end{document}